\documentclass[english,structabstract]{aa}
\usepackage{mathptmx}
\usepackage[T1]{fontenc}
\usepackage{txfonts}
\usepackage{babel}
\usepackage{array}
\usepackage{textcomp}
\usepackage{multirow}
\usepackage{amssymb}
\usepackage{amsfonts}

\usepackage{graphicx}
\usepackage{natbib}
\bibpunct{(}{)}{;}{a}{}{,} 

\bibpunct{(}{)}{;}{a}{}{,} 
\makeatother

\begin{document}
    
\title{Searching for tidal tails around $\omega$ Centauri using RR Lyrae Stars}

\subtitle{}

 \author{J. G. Fern\'andez-Trincado\inst{1,2,3}, 
 A. K. Vivas\inst{4},
 C. E. Mateu\inst{2,5},
 R. Zinn\inst{6},
 A. C. Robin\inst{1},
 O. Valenzuela\inst{7},
 E. Moreno\inst{7}
 \and 
 B. Pichardo\inst{7}
             }

 \institute{Institute Utinam, CNRS UMR6213, Universit\'e de Franche-Comt\'e, OSU THETA de Franche-Comt\'e-Bourgogne, Besan\c{c}on, France.\\
 \email{jfernandez@obs-besancon.fr}
 \and
 Centro de Investigaciones de Astronom\'ia, AP 264, M\'erida 5101-A, Venezuela.
 \and
 Postgrado en F\'isica Fundamental, Universidad de Los Andes, M\'erida, 5101, Venezuela.
 \and
 Cerro Tololo Interamerican Observatory, Casilla 603, La Serena, Chile.
 \and
 Instituto de Astronom\'ia, Universidad Nacional Aut\'onoma de M\'exico, Apdo. Postal 877, 22860 Ensenada, Baja California, Mexico.
  \and
 Department of Astronomy, Yale University, PO Box 20811, New Haven, CT 06520-8101, USA.
 \and
 Instituto de Astronom\'ia, Universidad Nacional Aut\'onoma de M\'exico, Apdo. Postal 70264, M\'exico D.F., 04510, Mexico.\\
 }

 \authorrunning{J. G. Fern\'andez-Trincado et al.}            


\date{Received 01/09/2014; accepted 07/11/2014}

\abstract{We present a survey for RR Lyrae stars
in an area of 50 deg$^2$ around the globular cluster $\omega$ Centauri, aimed to detect debris
material from the alleged progenitor galaxy of the cluster.  
We detected 48 RR Lyrae stars of which only 11 have been
previously reported. Ten among the eleven previously known stars were found inside the tidal radius
of the cluster. The rest were located outside the tidal radius up to distances of $\sim 6$ degrees
from the center of the cluster. Several of those stars are located at distances similar
to that of $\omega$ Centauri.
We investigated the probability that those stars may have been stripped off the cluster by
studying their properties (mean periods), calculating the expected halo/thick disk 
population of RR Lyrae stars in this part of the sky, analyzing the radial velocity of a sub-sample of the RR Lyrae 
stars, and finally, studying the probable orbits of this sub-sample around the Galaxy. 
None of these investigations
support the scenario that there is significant tidal debris around $\omega$ Centauri, 
confirming previous 
studies in the region. It is puzzling that tidal debris have been found elsewhere but not
near the cluster itself.}


\maketitle

{}

\section{Introduction}
\label{Sec:Introduction}
The globular cluster $\omega$ Centauri is the most massive
\citep{Meylan1987, Meylan1995, merr1997}  among the 157 
globular clusters known around the
Milky Way. It has several unusual properties that have led to the proposal that the
cluster is the remaining core of a dwarf galaxy destroyed due to gravitational interaction with the Milky Way
\citep{bekki2003,Mizutani+2003, Ideta+2004, Tsuchiya+2004, Majewski+2012}.  
Some of these unusual properties are: (\emph{i}) a retrograde, low inclination orbit around  
the Milky Way \citep{din99}, (\emph{ii}) an overall rapid rotation of 7.9 km s$^{-1}$\citep{merr1997}, 
making it one of the most flattened galactic globular clusters \citep{wh1987}, (\emph{iii}) a 
color-magnitude diagram showing a complex stellar population, with a wide range of metallicities,
two main sequences \citep{bed2004,sollima2007}, three red giant branches, and an age spread of 
$\sim$3-6 Gyr between the 
metal-poor and the most metal-rich populations \citep{Hilker2004,sollima2005}, (\emph{iv}) a complex chemical
pattern \citep{king2012,gra2011,marino2012}, and (\emph{v}) a
high velocity dispersion measured toward the center of the cluster, which has been interpreted as an
indication of having an intermediate mass black hole \citep{Noyola2010, 
Anderson2010, Miocchi2010, jalali2012}.
It has been proposed that $\omega$ Centauri may be
the equivalent to the M54 $+$ Sagittarius dwarf system but, in the case of $\omega$ Centauri,
the progenitor galaxy must have been already completely destroyed by now
\citep{cae10}. \citet{bekki2003} proposed a self-consistent dynamical model in which $\omega$ Centauri is the nucleus 
of a nucleated dwarf galaxy that was tidally destroyed when it merged with the first generation of the Galactic thin disk. \\ 

The search for remains of the alleged progenitor has not been without controversy. Although
\citet{leon2000} found significant
tidal tails coming out of the globular cluster $\omega$ Centauri, other work give opposite results.
\citet{leon2000} used wide-field multi-color images to do star counts around the cluster. They found
$\sim 7000$ stars localized outside of the
tidal radius of $r_t=45$ arcmin \citep{Trager1995} located along two tidal tails coming from the cluster 
from opposite directions, and aligned with the tidal field gradient, suggesting they are the
result of a collision with the Galactic disk.  Their results, however, may have been affected by 
reddening which may be high and variable around the cluster \citep{Law2003}. \\

On the other hand, \citet{dac08} made an extensive spectroscopic
survey of 4105 stars of the lower red giant branch in the vicinity of the cluster 
(a region of 2.4 $\times$ 3.9 deg$^2$). Only six of those red giant branch candidates had a velocity consistent with the 
radial velocity of cluster of ($+$232.2$\pm$0.7) km s$^{-1}$ \citep{din99}. 
\citet{dac08} concluded that these stars represent less than 1\% of the mass present in the cluster 
and hence, they do not provide a significant evidence of an extra-tidal population associated with $\omega$ Centauri. This would have 
been expected if most of the tidal stripping in the progenitor galaxy occurred a long time ago, which is the interpretation given by those 
authors to their results. \\

Interestingly, the search for $\omega$ Centauri debris seems to have been more successful in the
solar neighborhood by the recognition that stars in the Kapteyn group have kinematics and 
chemical abundance patterns similar to $\omega$ Centauri \citep{Wylie2010}. 
The chemical pattern was also key for the association of several red giants with retrograde orbits
studied by \citet{Majewski+2012}. Indeed, these authors suggest that $\omega$ Centauri is responsible for most
of the red giants in retrograde orbits in the inner halo.\\

Besides these successful identification of debris, one would like to, ideally, 
trace debris along other parts of the orbit as well, and most especially near
the cluster itself in 
order to understand better the origin of $\omega$ Centauri.
Since $\omega$ Centauri has a rich population of RR Lyrae stars \citep{kaluzny2004,delpri06,cacciari2006, weldra2007}, as
all satellite galaxies of the Milky Way do \citep[see for example,][]{vivas2006}, it is 
expected that any tidal debris from $\omega$ Centauri would also contain
this type of star. The use of RR Lyrae stars as tracers of debris around the cluster has several advantages. They are 
bright stars which are relatively easy to spot at different distances because of their variability 
properties. They are standard candles and hence we can identify possible debris as stars at the same 
distance as the cluster. Finally,  they are an old population and hence we expect no contamination by 
the thin disk (although we still have to deal with thick disk contamination). Extensive previous work 
have demonstrated that RR Lyrae stars are excellent tracers of substructures in the Halo 
\citep[][among others]{vivas2006,watkins2009,drake2013}. We present here a survey for RR Lyrae stars in a region of
$\sim 50$ deg$^2$ around $\omega$ Centauri. Preliminary results of this survey were presented in \citet{Fernandez-Trincado2013}.\\

In Section 2, we describe the observations. The methods for selecting variable stars of the RR Lyrae type are 
presented in Section 3. The properties, distances and spatial distribution of the RR Lyrae stars detected in this work 
are discussed in the Section 4. Section 5 analyzes the likelihood that these RR Lyrae stars are part of debris from the
destroyed progenitor galaxy of $\omega$ Cen. Finally, 
conclusions are presented in Section 6.\\

\section{Observations}
\label{Sec:Data}

The techniques used for this survey are similar to the ones used extensively by our group in studies of RR Lyrae stars in the galactic halo and the 
Canis Major over-density with the QUEST\footnote{Quasar Equatorial Survey Team} camera \citep{vivas2004,mat2009}.
The photometric survey was carried out using the QUEST camera at the 1.0m J\"urgen Stock telescope (1.5m Schmidt Camera) 
at the National Astronomical Observatory of Llano del Hato, Venezuela.  The QUEST camera is 
a mosaic of 16 CCDs, with a field of view of 2.3$\times$2.5 deg$^2$. Each detector has 
2048$\times$2048 pixel of 15 $\mu$m, resulting in an angular resolution of 1 arcsec/pix \citep{baltay2002}. \\

For the present survey, half of the camera (8 CCDs) was covered with V filters and the other half with I filters.  
Although the QUEST camera was designed to work more efficiently in drift scan mode near the equator,
the high declination of $\omega$ Centauri required to work in a classical point-and-stare mode. An 
inconvenience of this method is that it is not possible to cover the whole focal plane with the same filter. Hence, 
appropriate offsets have to be made in order to have uniform covering of the same area of the sky in more 
than one filter (at any pointing, half of the camera is observing
through one filter and the other half with another). Due to adverse weather conditions not all of the survey area
was observed in both bands. \\
 
\begin{center}
\begin{figure}
\includegraphics[width=100mm]{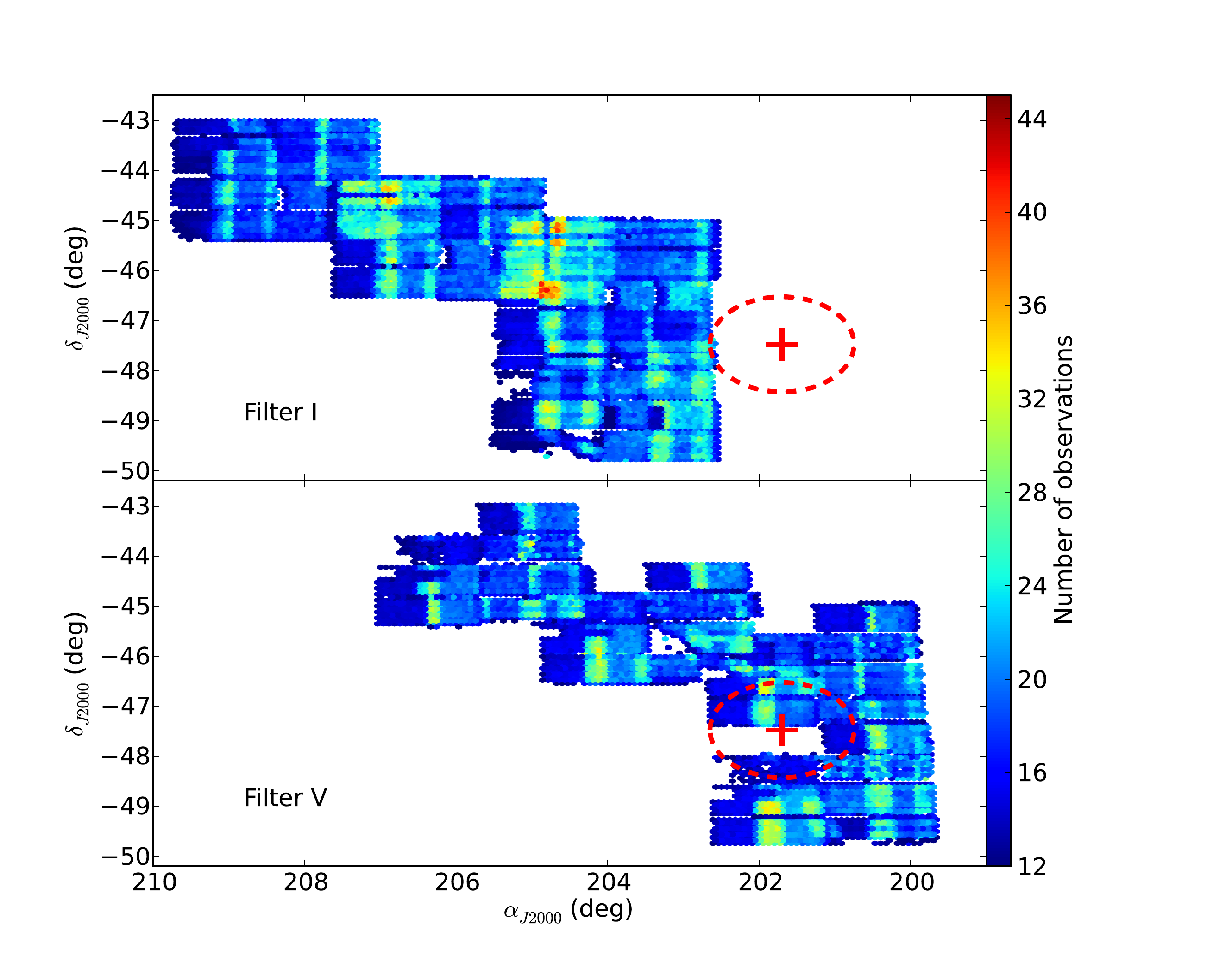}
 \caption{Density of observations around of $\omega$ Centauri in the V (bottom) and I (top) band: the color scale is proportional to the number of observations in each band.
  The large '+' symbol and dashed ellipse represent the center and tidal radius ($r_t=57$ arcmin)
 of the cluster, respectively.}
 \label{density}
\end{figure}
\end{center}

Multi-epoch observations were obtained  for fields around $\omega$ Centauri during 18 nights between 
the years 2010 and 2011. Some nights fields were observed more than once,
separated by at least 1 hour.
The total area covered with multi-epoch observations (either in only one photometric band or both)
was $\sim 50$ deg$^2$. 
We used exposure times of 60s and 90s. Observations of $\omega$ Centauri from Llano
del Hato (at a latitude of $+8\textordmasculine 47\arcmin$) are challenging since the cluster never gets high in the 
sky. We, however, avoided observations with airmass $> 2$. Average seeing was around 3 arcsec 
which was partly a consequence of observing at very high airmasses.
Figure ~\ref{density} shows the density of the observations in each band in the area of the survey.
The fields were chosen to cover a section of the orbit of $\omega$ Centauri, in the 
opposite direction to the movement of the cluster, 
according to its proper motion \citep{din99}.\\

For the data processing 
we used the standard IRAF tasks for overscan, bias and flat fielding corrections. 
Flat Fielding was made by constructing synthetic flats from a large number of sky observations since this procedure gave
significantly better results that using dome flats.
Aperture photometry was
performed using the APPHOT task of IRAF. 
The use of aperture photometry is justified because our main interest is 
the region around the cluster and not the cluster itself. Away from the center of the cluster, the density 
of stars is low ($\sim 4,500$ stars per CCD). \\

Astrometry was done using the program CM1 \citep{stock1981} which calculates the transformation matrix based on coordinates from the UCAC4 catalog
\citep{zacha2013}. The precision of the astrometric solutions was of the order of 0.2 arcsec.\\

 All magnitudes were normalized to a reference catalog following
the methodology used by \citet{vivas2004}. The normalization 
was performed independently in each CCD by using 500 to 1000 stars in 
each image. An ensemble clipped mean of the differences between magnitudes in each image and
the reference image was calculated and added to all individual magnitudes in that image. 
The error added due to this 
zero point normalization was typically  $<0.07$ mag.\\

Then, instrumental magnitudes in the reference images were calibrated using zero point corrections estimated by matching 
stars with two different catalogs: APASS\footnote{The American Association of Variable Star Observers Photometric All-Sky Survey} (Henden et al. 2012), and DENIS\footnote{Deep Near Infrared 
Survey of the Southern Sky} (Epchtein et al. 1997), 
for stars observed in the V and I bands, respectively.\\

Saturation and limiting magnitudes in our catalog are respectively 11 and 19 mag in I, and 12.5 and 20 mag in V. 
Our final catalog has 456,539 stars.\\

\begin{center}
\begin{figure}
\includegraphics[width=100mm]{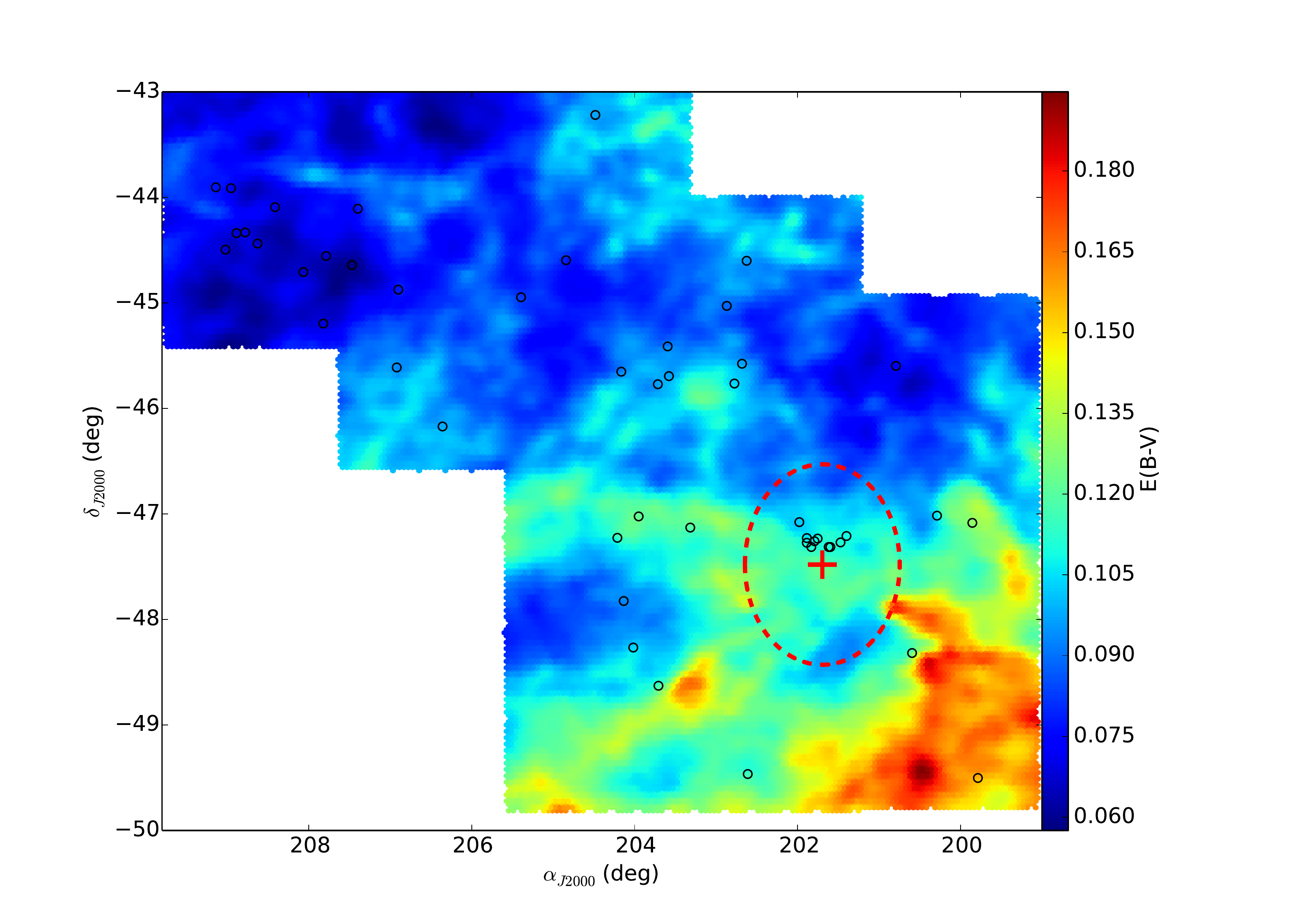}
 \caption{ Reddening ($E{(B-V)}$) map around $\omega$ Centauri from \citet{Schlafly+2011}. 
 RR Lyrae stars detected in this work are represented by open circles. 
 The center of the cluster and its tidal radius are represented by the red '+' symbol and dashed ellipse, respectively.}
\label{extinction}
\end{figure}
\end{center}

$\omega$ Centauri is located at a galactic latitude of $b=14.97\textordmasculine$ \citep{harr1996} which is low enough
to expect large variations in extinction across the region. Dust maps from \citet{schlegel1998}, with the re-calibration proposed by
\citet{Schlafly+2011}, were used for correcting by interstellar extinction each individual star\footnote{\url{http://irsa.ipac.caltech.edu/applications/DUST/}}.  The standard deviation of the interstellar extinction along all the region 
is $\sigma_{A_V}=0.07$ mag and $\sigma_{A_I}=0.04$ mag. A map of the color excess, E(B-V), over the survey region
is shown in the Figure ~\ref{extinction}.\\

\section{Search of RR Lyrae stars around $\omega$ Centauri}

\subsection{Selection of variable stars in the Color-Magnitude Diagram}

\begin{figure}
\includegraphics[width=105mm]{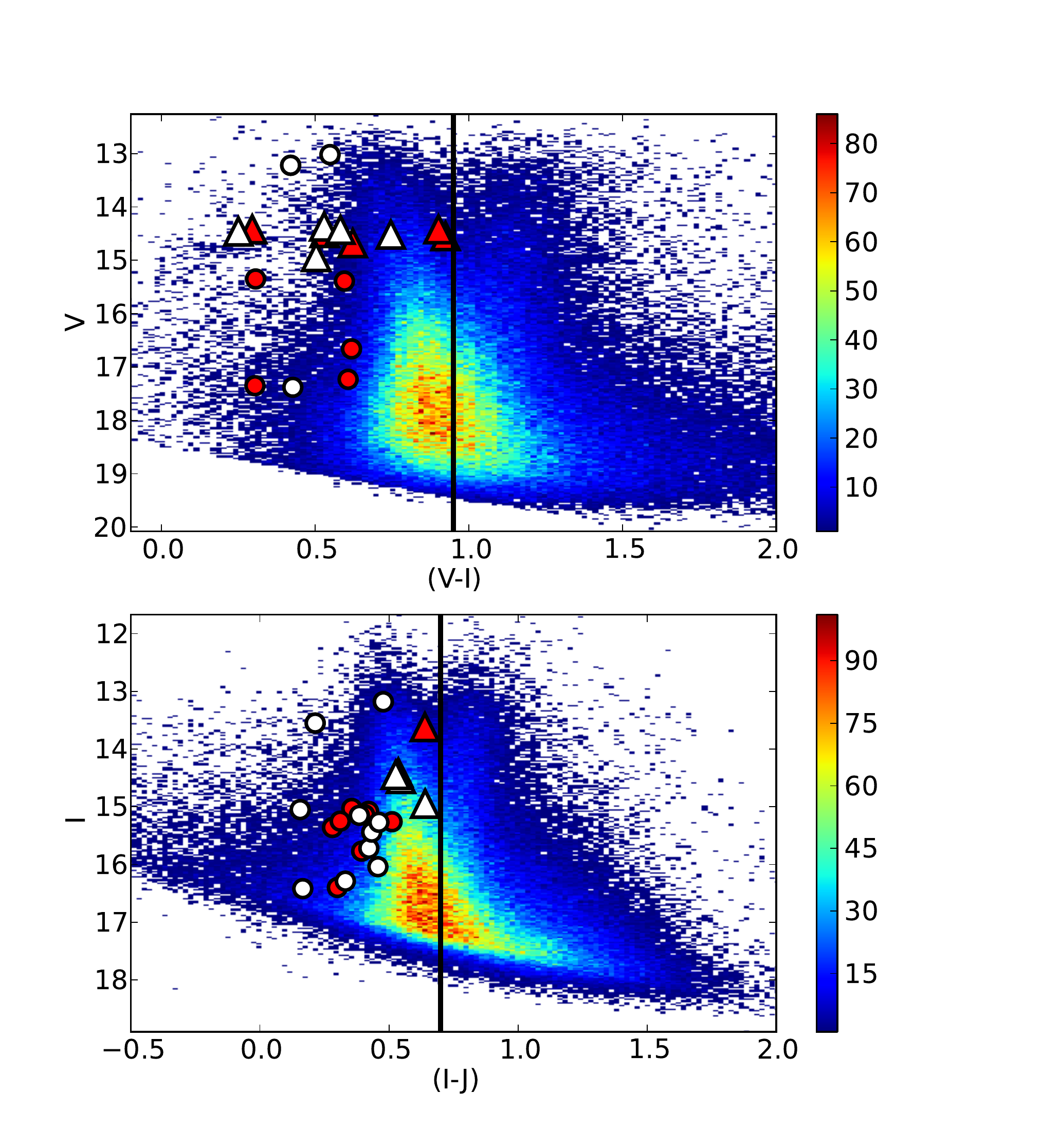}
\caption{Hess diagrams for two different regions of the survey. The top panel shows a region which was observed
in both V and I bands in our survey. The region shown in the bottom panel was observed only in I and we used 2MASS J-band to 
construct a color-magnitude diagram. 
Black vertical lines indicate the color cut for selecting RR Lyrae star candidates (however, the most interesting region was explored with no color cut).
The triangle symbols indicate the confirmed RR Lyrae stars having a brightness similar to the Horizontal Branch of $\omega$ 
Centauri (see \S 4.2). Circles correspond to other RR Lyrae stars in the field of view (stars in the foreground or background of the cluster). 
RR Lyrae stars of the type $ab$ and $c$ are presented
as red  and white symbols, respectively. The color scale is 
proportional to the number of stars in each bin.}
 \label{CMD}
\end{figure}

RR Lyrae stars are horizontal branch stars with spectral types A and F. A simple color cut would 
allow to significantly reduce 
the number of candidates to RR Lyrae stars. However, as explained in the last section not all of the 
survey area was observed in two photometric bands.
To overcome this difficulty, we cross-matched our catalog with 2MASS 
\citep{skrutskie2006} to obtain either the $(V-J)$ or $(I-J)$ color in those regions observed  in only one band.
Magnitudes from 2MASS are single-epoch and RR Lyrae stars are 
expected to change color (change $T_{\rm eff}$) during the
pulsation cycle. In addition, mean magnitudes may not be accurate if the lightcurves
are poorly sampled. Thus, we searched for RR Lyrae stars with no color cut at all in the range of magnitudes of most
interest for this work, $13.0 < V < 16.5$ (that is, around the magnitude of the horizontal branch of $\omega$ Cen).
For the rest of the sample we applied the following color cut: 
$(V-I) \leq 0.95$ or $(I-J) \leq 0.70$ (Figure~\ref{CMD}). \\  

The next step toward identifying RR Lyraes was to detect
variable stars in our time-series catalog. We calculated the Pearson 
distribution $\chi^2$ (Eq.~\ref{chi}) for all stars that passed our color cut and selected 
those ones whose probability is $P(\chi^2) < 0.01$, which
corresponds to a 1\% probability of the magnitude distribution being due to the
observational errors:\\

\begin{equation}
\label{chi} \chi^2=\sum^{N}_{i=1}\frac{(m_i-\langle m \rangle)^2}{\nu \sigma^2_i}
\end{equation}

\noindent
where, $m_i$ ...., $m_{N}$ are the individual
magnitudes with observational errors $\sigma_i$,
 $\langle m \rangle$ is the mean magnitude, and $\nu=N-1$ is the degrees of freedom ($N$ is the number
 of observations by star).\\

Before calculating the $\chi^2$ probability we eliminated 
measurements that were potentially affected by cosmic
rays or bad pixels by deleting any point whose magnitude was more than 4$\sigma$
away from the average magnitude of the star. This step helps to eliminate spurious variability.\\

\subsection{Selection of RR Lyrae stars}

\begin{figure*}
\begin{center}
 \includegraphics[width=164mm,height=205mm]{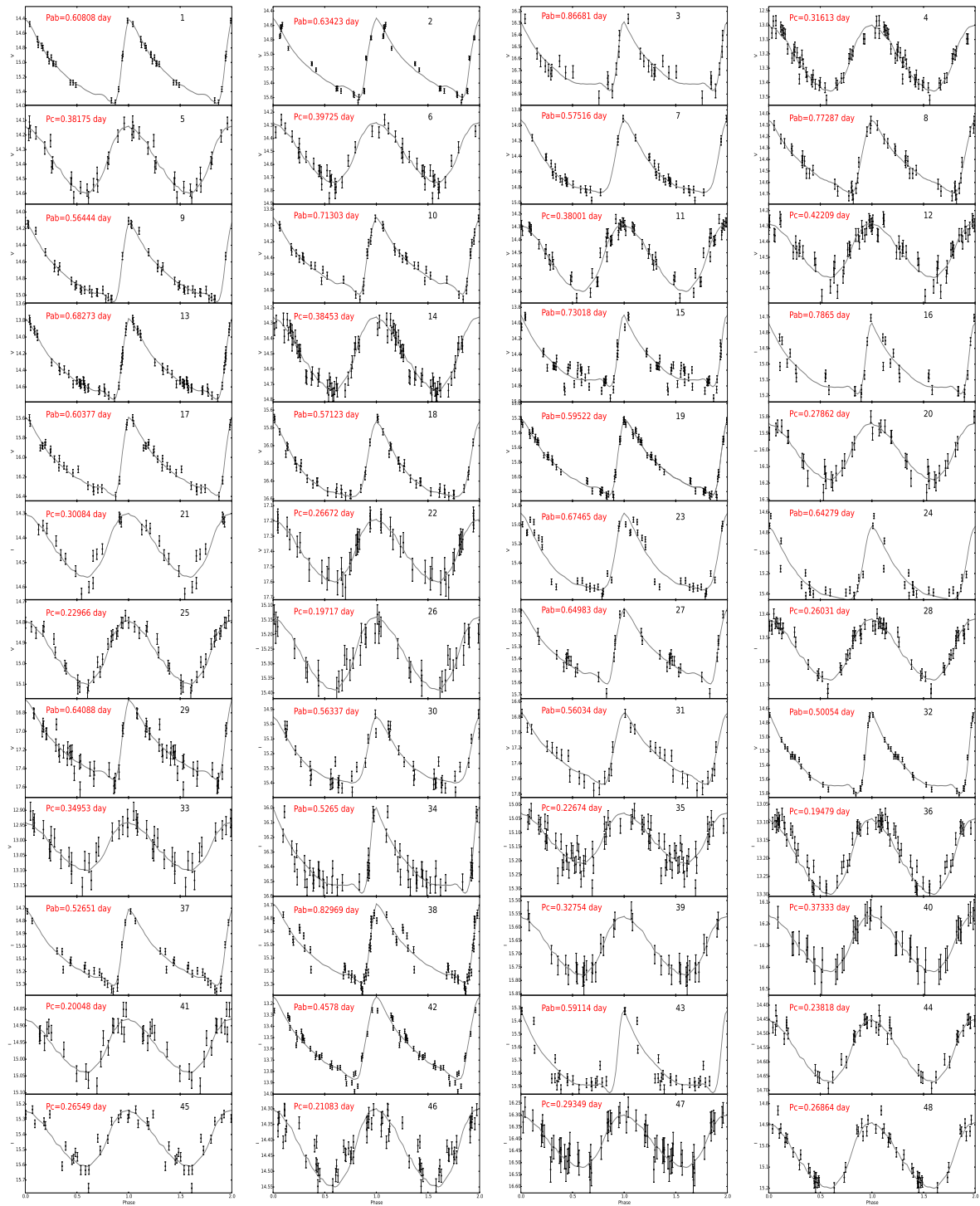}
 \caption{Light curves for RR Lyrae stars. Solid lines are the best template fitted to each light curve.}
 \label{lightcurves}
 \end{center}
 \end{figure*}

Best possible periods in the range 0.2-1.2 days were determined for all variable candidates using the \citet{laf1965} algorithm 
\citep[see also][]{vivas2004}. Phased lightcurves were visually examined.  
RR Lyrae stars were finally identified based on their amplitude, period and the 
shape of the light curves. Forty-seven RR Lyrae stars (25 RRab and 23 RRc) were detected, 37 of which 
are new discoveries. Periods, amplitudes and ephemerids were refined by fitting templates of 
lightcurves of RR Lyrae stars to the data points, following the procedure described in \citet{vivas2008}.
Lightcurves are shown in Figure~\ref{lightcurves}.\\

Distances were calculated by assuming the absolute magnitude relationships given by \citet{catelan04}: 

\begin{equation}
\label{MI} M_I= 0.47 - 1.132logP + 0.205logZ,\\
\end{equation}

\begin{equation}
\label{MV} M_V= 2.288 + 0.882logZ + 0.108logZ^2, \\
\end{equation}

\noindent
where

\begin{equation}
logZ= [Fe/H] + log(0.638\times10^{[\alpha/Fe]}+0.362) - 1.765, \\
\end{equation}

 \citet{Sollima+2006} measured the metal abundance of 74 RR Lyrae stars in $\omega$ Cen, obtaining a mean value
of [Fe/H]$=-1.7$ dex. On the other hand, following \citet{delpri06}, we adopted a value of $0.477$ for the
$\alpha$-element abundance. Table~\ref{table1} contains all the relevant information for the 48 RR Lyrae stars: 
column 1 indicates the star ID; columns 2 and 3 correspond to the right ascension and declination, 
respectively, 
whereas column 4 shows the number of times each star was observed. Columns 5-8 list the 
type of RR Lyrae star, our derived period and amplitude, and the heliocentric Julian day at maximum 
light. Columns 9 and 10 show the mean V and I magnitudes. The average E(B-V) 
in an area with 15 arcmin radius around each star are given in column 11. 
The heliocentric distance and the angular distance (in degrees) from the center
of $\omega$ Centauri are 
shown in columns 12 and 13, respectively. Radial velocities for a sub-sample of the stars are reported in column 14. Finally, the ID number
in the May 2014 version of the catalog of variable stars in globular clusters 
by \citet[][C01]{clement01} is given in column 15.\\  

For several type c stars we found that one of 1-day aliases produced phased light curves 
as good as the one with the main 
period found for those stars. Based in our data there is no way to determine which one is the true 
period and which one is the spurious period. 
Table~\ref{table1} contains double entries for the periods in those cases.\\

We estimated our completeness by producing a set of simulated light curves of RR Lyrae stars
with the same temporal sampling and photometric
errors as our data. The artificial light curves were then analyzed with the same tools we used
for our data. At the typical magnitude range of RR Lyrae stars in $\omega$ Centauri, 
we were able to successfully recover the periods of type ab stars in $\gtrsim 90\%$
of the cases if the number of epochs (N) was larger than 20. The completeness drops to 85\% for
N=17 and 56\% for N=12. The completeness of type c stars is lower; it is around 70\% for
the best sampled regions ($N>17$), dropping to 40\% for N=12.\\

\begin{table*}
\setlength{\tabcolsep}{1.35mm}  
\begin{tiny}
\caption{RR Lyrae Stars}
\begin{tabular}{ccccccccccccccccccccccccc}
\hline
ID & $\alpha_{2000}$       & $\delta_{2000}$      & $N_{obs}$ & Type & P & Amp   & HJD &  $\langle V \rangle$   &  $\langle I \rangle$  & E(B-V)  & $D$   & $\theta$ & RV            & ID(C01) \\
   &  (deg)   &  (deg)  &           &      & (day)   & (mag)   &     (day)     & (mag)  & (mag) &  mag    & (kpc) & (deg) & (km s$^{-1}$) &       \\
\hline
  1      &    199.787506 &    -49.50225  &     22    &     ab    &     0.60808 &     1.15  &     2455706.66692 &     15.06$\pm$0.02 &     -              &      0.155  &      6.3   $\pm$1.0  &     2.4   &     114$\pm$19 &     -     \\
  2      &    199.855804 &    -47.08439  &     24    &     ab    &     0.63423 &     1.10  &     2455309.71655 &     15.11$\pm$0.02 &     -              &      0.130  &      6.6   $\pm$1.0  &     1.3   &     293$\pm$20 &     -     \\
  3      &    200.289261 &    -47.016312 &     17    &     ab    &     0.86681 &     0.58  &     2455309.73527 &     16.66$\pm$0.16 &     -              &      0.107  &      14.0  $\pm$3.0  &     1.1   &     -           &     -     \\
  4      &    200.595367 &    -48.31789  &     40    &     c     &     0.31613 &     0.46  &     2455324.63072 &     13.22$\pm$0.03 &     -              &      0.130  &      2.8   $\pm$0.4  &     1.1   &     -           &     175   \\
  5      &    200.793121 &    -45.59691  &     21    &     c     &     0.38175 &     0.46  &     2455597.8221  &     14.37$\pm$0.04 &     -              &      0.074  &      5.1   $\pm$0.8  &     2.0   &     123$\pm$ 10&   -       \\
  6      &    201.40033  &    -47.20893  &     25    &     c     &     0.39725 &     0.46  &     2455324.68931 &     14.52$\pm$0.03 &     -              &      0.109  &      5.2   $\pm$0.8  &     0.3   &     -           &     160   \\
  7      &    201.473541 &    -47.2696   &     25    &     ab    &     0.57516 &     0.94  &     2455655.76307 &     14.49$\pm$0.03 &     -              &      0.111  &      5.1   $\pm$0.8  &     0.3   &     280$\pm$53 &     73    \\
  8      &    201.597946 &    -47.31337  &     25    &     ab    &     0.77287 &     0.63  &     2455655.76505 &     14.42$\pm$0.03 &     -              &      0.111  &      5.0   $\pm$0.8  &     0.2   &     -           &     54    \\
  9      &    201.619019 &    -47.31311  &     25    &     ab    &     0.56444 &     0.95  &     2455706.66799 &     14.68$\pm$0.03 &     -              &      0.111  &      5.6   $\pm$0.8  &     0.2   &     -           &     67    \\
  10     &    201.754318 &    -47.233349 &     25    &     ab    &     0.71303 &     0.95  &     2455596.91557 &     14.42$\pm$0.27 &     -              &      0.109  &      5.0   $\pm$1.3  &     0.2   &     -           &     7     \\
  11     &    201.792587 &    -47.258259 &     35    &     c     &     0.38001 &     0.52  &     2455706.69839 &     14.46$\pm$0.17 &     -              &      0.108  &      5.1   $\pm$1.1  &     0.2   &     -           &     36    \\
  12     &    201.832062 &    -47.313049 &     37    &     c     &     0.42209 &     0.35  &     2455597.79094 &     14.43$\pm$0.14 &     -              &      0.111  &      5.0   $\pm$1.0  &     0.2   &     -           &     75    \\
  13     &    201.886887 &    -47.22865  &     41    &     ab    &     0.68273 &     0.97  &     2455309.74263 &     14.33$\pm$0.03 &     -              &      0.106  &      4.8   $\pm$0.7  &     0.3   &     216$\pm$ 19&   149     \\
  14     &    201.887589 &    -47.27296  &     41    &     c     &     0.38453 &     0.48  &     2455596.84259 &     14.50$\pm$0.03 &     -              &      0.109  &      5.2   $\pm$0.8  &     0.2   &     -           &     72    \\
  15     &    201.97934  &    -47.07737  &     38    &     ab    &     0.73018 &     0.91  &     2455597.88609 &     14.60$\pm$0.19 &     -              &      0.102  &      5.5   $\pm$1.2  &     0.4   &     -           &     172   \\
  16     &    202.612198 &    -49.46468  &     17    &     ab    &     0.7865  &     0.45  &     2455323.76461 &     -               &  15.03$\pm$0.03   &      0.114  &      9.6   $\pm$0.4  &     2.1   &     -           &     -     \\
  17     &    202.62558  &    -44.60067  &     25    &     ab    &     0.60377 &     0.80  &     2455309.68646 &     16.05$\pm$0.03 &    -               &      0.095  &      10.8  $\pm$1.6  &     3.0   &     -           &     -     \\
  18     &    202.68248  &    -45.576481 &     27    &     ab    &     0.57123 &     0.85  &     2455706.63748 &     16.26$\pm$0.04 &    -               &      0.097  &      11.8  $\pm$1.8  &     2.0   &     -           &     -     \\
  19     &    202.775803 &    -45.764069 &     41    &     ab    &     0.59522 &     1.04  &     2455321.75328 &     15.76$\pm$0.03 &     -              &      0.103  &      9.3   $\pm$1.4  &     2.0    &     -           &     -     \\
  20     &    202.869476 &    -45.028179 &     21    &     c     &     0.27862 &     0.34  &     2455596.85697 &     -               &  16.04$\pm$0.14   &      0.088  &      12.2  $\pm$1.1  &     2.6   &     -           &     -     \\
  21     &    203.317429 &    -47.12904  &     16    &     c     &     0.30084 &     0.26  &     2455655.75493 &     -               &  14.45$\pm$0.10   &      0.120  &      5.9   $\pm$0.4  &     1.2   &     -           &     -     \\
  22     &    203.579575 &    -45.693771 &     25    &     c     &     0.26672 &     0.41  &     2455323.70735 &     17.38$\pm$0.15 &    -               &      0.097  &      19.8  $\pm$4.1  &     2.2   &     -           &     -     \\
  23     &    203.595047 &    -45.412022 &     25    &     ab    &     0.67465 &     0.80  &     2455597.82386 &     15.39$\pm$0.02 &    -               &      0.092  &      8.0   $\pm$1.2  &     2.4   &     -           &     -     \\
  24     &    203.707916 &    -48.628139 &     21    &     ab    &     0.64279 &     0.72  &     2455309.67323 &     -               &  15.25$\pm$0.23   &      0.114  &     10.1   $\pm$1.3  &     1.8   &     -           &     -     \\
  25     &    203.715485 &    -45.770672 &     27    &     c     &     0.22966 &     0.31  &     2455706.68115 &     14.94$\pm$0.02 &    -               &      0.096  &      6.5   $\pm$0.9  &     2.2   &     -55$\pm$10 &     -     \\
  $26^a$ &    203.950912 &    -47.023251 &     21    &     c     &     0.19717 &     0.19  &     2455309.76025 &     -               &  15.27$\pm$0.08   &      0.122  &      7.8   $\pm$0.5  &     1.6   &     -           &     -     \\
   &            &            &    &      & 0.24183 &      &               &                &                &       &            &     &           &                                                                          \\
  27     &    204.017181 &    -48.265869 &     19    &     ab    &     0.64983 &     0.63  &     2455324.68881 &     -               &  15.36$\pm$0.17   &      0.104  &      10.7  $\pm$1.1  &     1.7   &     -           &     -     \\
  28     &    204.134323 &    -47.82508  &     35    &     c     &     0.26031 &     0.26  &     2455321.70224 &     -               &  13.55$\pm$0.02   &      0.094  &      3.8   $\pm$0.1  &     1.7   &     -           &     -     \\
  29     &    204.164261 &    -45.651611 &     32    &     ab    &     0.64088 &     0.86  &     2455597.7796  &     17.23$\pm$0.23 &    -               &      0.091  &      18.6  $\pm$4.5  &     2.5   &     -           &     -     \\
  30     &    204.212128 &    -47.2262   &     25    &     ab    &     0.56337 &     0.45  &     2455309.69697 &     -               &  15.26$\pm$0.16   &      0.115  &      9.8   $\pm$1.0  &     1.7   &     -           &     -     \\
  31     &    204.482224 &    -43.218788 &     17    &     ab    &     0.56034 &     0.92  &     2455324.67663 &     17.35$\pm$0.29 &    -               &      0.097  &      19.5  $\pm$5.3  &     4.7   &     -           &     -     \\
  32     &    204.84198  &    -44.595249 &     24    &     ab    &     0.50054 &     1.16  &     2455596.91508 &     15.35$\pm$0.25 &    -               &      0.079  &      8.0   $\pm$2.0  &     3.6   &     -           &     -     \\
  33     &    205.394684 &    -44.94603  &     24    &     c     &     0.34953 &     0.16  &     2455655.74781 &     13.02$\pm$0.06 &    -               &      0.082  &      2.7   $\pm$0.4  &     3.6   &     -           &     -     \\
  34     &    206.357758 &    -46.170818 &     31    &     ab    &     0.5265  &     0.58  &     2455309.72722 &     -               &  16.40$\pm$0.15   &      0.097  &      16.6  $\pm$1.6  &     3.4   &     -           &     -     \\
  35     &    206.900879 &    -44.87455  &     40    &     c     &     0.22674 &     0.18  &     2455706.62415 &     -               &  15.15$\pm$0.07   &      0.078  &      7.8   $\pm$0.4  &     4.4   &     -           &     -     \\
  $36^a$ &    206.92006  &    -45.61159  &     37    &     c     &     0.19479 &     0.21  &     2455324.70204 &     -               &  13.18$\pm$0.07   &      0.094  &      3.0   $\pm$0.2  &     4.0   &     -           &     -     \\
   &            &            &    &      & 0.24209 &      &               &                &                &       &              &     &           &      \\
  37     &    207.397919 &    -44.10767  &     21    &     ab    &     0.52651 &     0.60  &     2455323.71549 &     -               &  15.14$\pm$0.18   &      0.077  &     9.4   $\pm$1.0 &     5.2   &     -           &     -     \\
  38     &    207.469147 &    -44.640949 &     39    &     ab    &     0.82969 &     0.60  &     2455596.91604 &     -               &  15.08$\pm$0.19   &      0.059  &     10.3   $\pm$1.1&     4.9   &     -           &     -     \\
  39     &    207.786301 &    -44.555771 &     21    &     c     &     0.32754 &     0.22  &     2455309.76984 &     -               &  15.72$\pm$0.08   &      0.066  &      11.1 $\pm$0.7 &     5.1   &     -           &     -     \\
  40     &    207.820847 &    -45.19574  &     23    &     c     &     0.37333 &     0.26  &     2455309.79996 &     -               &  16.29$\pm$0.10   &      0.068  &      14.9 $\pm$1.0 &     4.8   &     -           &     -     \\
  41     &    208.065308 &    -44.707298 &     22    &     c     &     0.20048 &     0.16  &     2455309.75057 &     -               &  14.95$\pm$0.06   &      0.066  &      7.0  $\pm$0.4 &     5.2   &     -           &     -     \\
  42     &    208.413483 &    -44.092201 &     36    &     ab    &     0.4578  &     0.73  &     2455655.76003 &     -               &  13.62$\pm$0.20   &      0.070  &      4.6  $\pm$0.5 &     5.8   &     -           &     -     \\
  43     &    208.628021 &    -44.437511 &     23    &     ab    &     0.59114 &     0.63  &     2455309.72658 &     -               &  15.77$\pm$0.18   &      0.066  &      13.0 $\pm$1.4 &     5.7   &     -           &     -     \\
  44     &    208.780243 &    -44.331718 &     24    &     c     &     0.23818 &     0.22  &     2455593.84126 &     -               &  14.52$\pm$0.08   &      0.069  &      6.0  $\pm$0.4 &     5.8   &     -           &     -     \\
  45     &    208.886917 &    -44.338421 &     23    &     c     &     0.26549 &     0.37  &     2455590.83373 &     -               &  15.44$\pm$0.15   &      0.069  &      9.3  $\pm$0.9 &     5.9   &     -           &     -     \\
  $46^a$ &    208.95462  &    -43.91272  &     36    &     c     &     0.21083 &     0.17  &     2455597.80525 &     -               &  14.41$\pm$0.07   &      0.072  &      5.5  $\pm$0.3 &     6.2   &     -           &     -     \\
   &            &            &    &      & 0.26712 &      &               &                &                &       &              &     &           &      \\
  47     &    209.022217 &    -44.495121 &     34    &     c     &     0.29349 &     0.22  &     2455597.81265 &     -               &  16.42$\pm$0.08   &      0.068  &      15.0 $\pm$0.9  &     5.9   &     -           &     -     \\
  48     &    209.140732 &    -43.90432  &     30    &     c     &     0.26864 &     0.31  &     2455593.83375 &     -               &  15.05$\pm$0.12   &      0.077  &      7.8  $\pm$0.6  &     6.3   &     -           &     -     \\
\hline
\end{tabular}  \label{table1}
\tablefoot{$^a$ Two periods reported for these stars (see text).}
\end{tiny}
\end{table*}

\section{Properties of the RR Lyrae Stars}

\subsection{Previously known stars}

Eleven out of the 48 RR Lyrae stars detected in our survey were already reported in the literature, all of them as members of $\omega$ Centauri \citep{clement01}.
Ten of those stars are located between 10 and 25 arcmin from the center of $\omega$ Centauri; that is, well inside the tidal radius of the cluster (57 arcmin).
There is no doubt these are cluster members which were recovered by our survey.
The remaining star, \#4 \citep[or V175 in][]{clement01}, is in the outskirts of the cluster, at 66 arcmin from its center. The original source for this star goes back to
\citet{wilkens65} and no period or type of variable is reported.  From our data, this star seems to be a real periodic star with a period of 0.316 d and a
sinusoidal light curve. Although it resembles the properties of an RR Lyrae star of the type $c$, its mean magnitude is more than one
magnitude brighter than the horizontal branch of the cluster. Bright variables such as Anomalous Cepheids and Population II Cepheids (BL Her) have been found in $\omega$ Cen
\citep{Kaluzny+1997}. However, those stars have periods ranging from 0.5 to 2.3d which are significantly larger than the period of V175.
This seems to indicate that V175 is actually a foreground variable and not a real member of the cluster, although
a measurement of the radial velocity would be needed to confirm it. \\ 

Our classification, periods and amplitudes for the 10 stars (6 of the type $ab$ and 4 of the type $c$) in the cluster agree quite well with the values
reported in the list of variable stars in $\omega$ Centauri catalog 
\citep[][May 2014 version]{clement01}
The average of the absolute value of the differences between our periods and the published ones 
is only $8\times10^{-4}$ days.  
Reassuringly, also the mean V magnitudes agree within 0.03 mags. These good agreements validate our methods
for the photometry and detection of variables.\\

We found no new variables within the tidal radius of the cluster. On the other hand, we did not recover all known variables in the cluster but this is due
to the fact that we intentionally left out of the survey most of the central part of the cluster. The reason is twofold: our objective is to study tidal debris
around the cluster and then there is no special interest in the cluster itself, and, on the other hand, the resolution and median seeing of our data ($\sim$3 arcsec) is
not adequate for crowded field photometry.\\ 
 
\subsection{Distances and Spatial Distribution}

  The distance to $\omega$ Centauri based on the average of the 10 RR Lyrae stars inside its tidal radius is ($5.15 \pm 0.23$) kpc. This value 
agrees very well with the distance of $5.4\pm0.7$ derived by \citet{weldra2007} using (optical) observations of 69 RR Lyrae stars in the cluster,
although it is somewhat short if compared with the value of $5.57 \pm 0.08$ kpc recently derived by \citet[][in prep]{Navarrete+2015} from IR lightcurves 
of a similar number of RR Lyrae stars.\\

\begin{figure}
\begin{center}
\includegraphics[width=90mm]{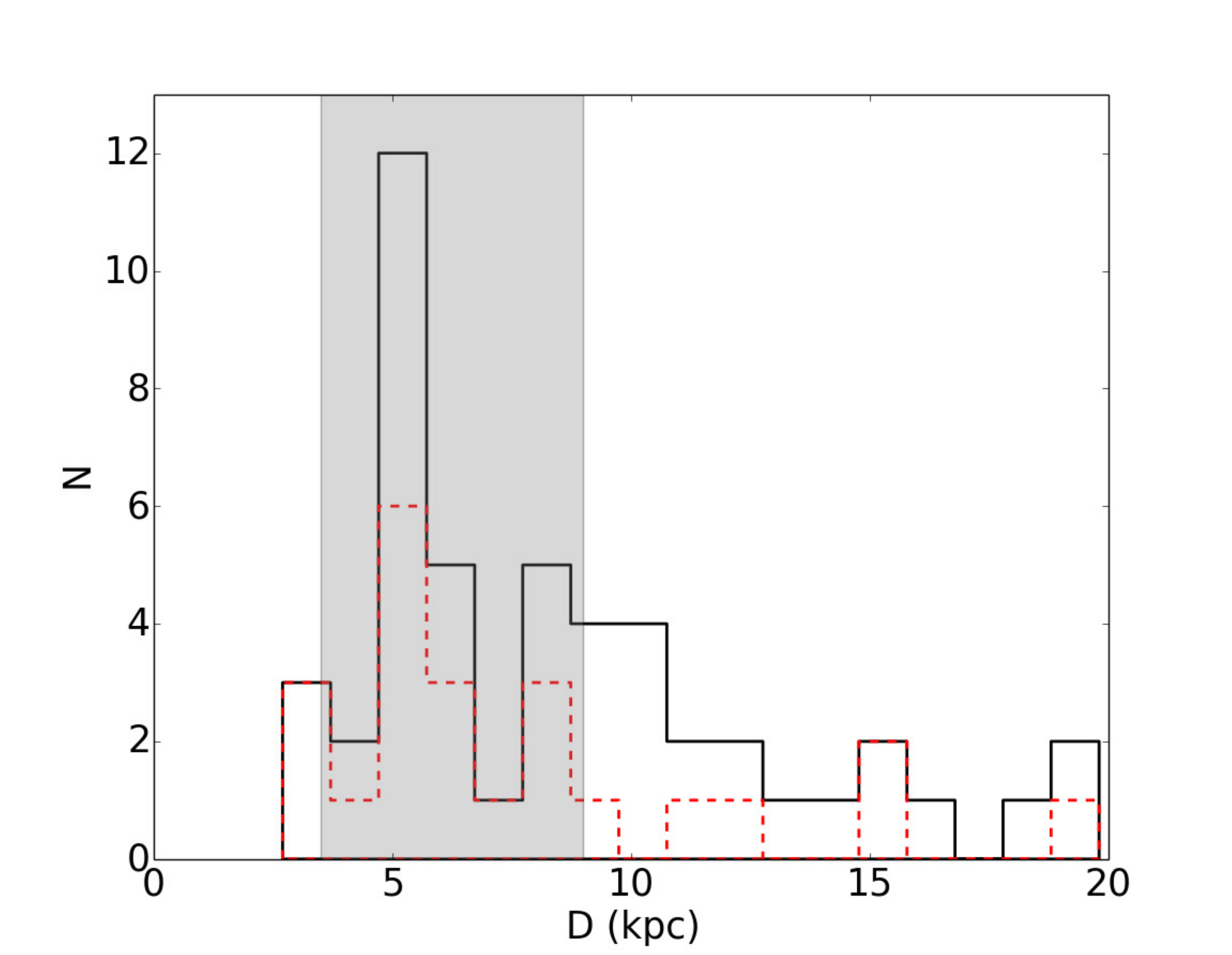}
\caption{Distribution of distances of the RR Lyrae stars (solid line histogram). The distribution of the 
type-c RR Lyrae stars is shown with a red dashed histogram. The shaded region
encloses the region of stars selected as candidates to tidal debris from the cluster.}
\label{distances}
\end{center}
\end{figure}

A histogram of the distribution of distances from the Sun for all the stars found in our survey can be seen in
Figure~\ref{distances}. The histogram shows a pronounced 
peak at about $D = $ 5.2 kpc which naturally corresponds to the 10 cluster RR Lyrae stars discussed above. 
We used this diagram to select other stars around the cluster having a similar distance. Those stars may have been tidally disrupted from the cluster.
The shaded region in the histogram encloses the region  ($3.5 \leq D \leq 9 $ kpc) 
of our candidates for tidal debris. Within these limits we found  15 RR Lyrae stars outside the tidal radius of the cluster (that is,
not counting the 10 cluster RR Lyrae stars).\\ 

\begin{figure}
\begin{center} 
\includegraphics[width=97mm]{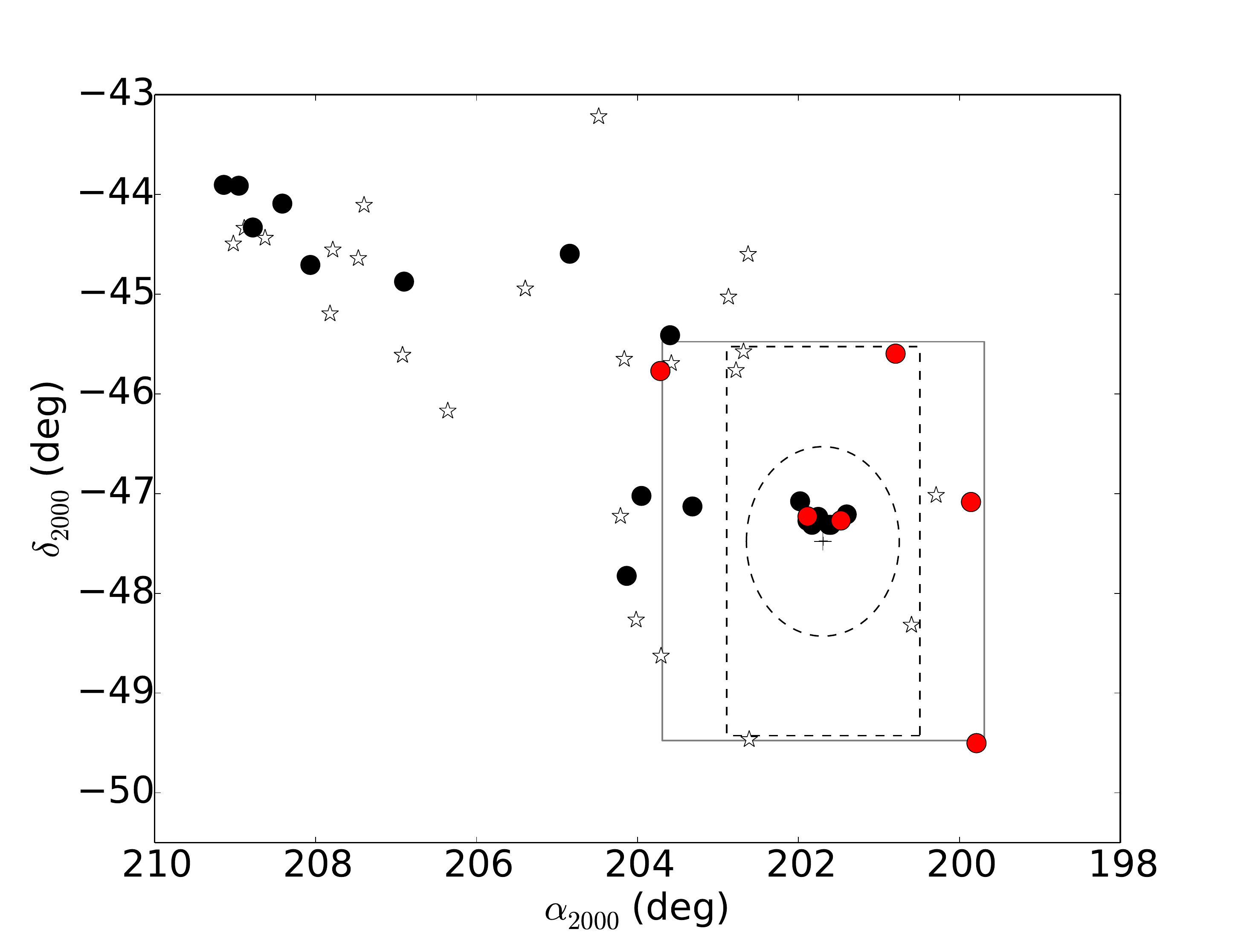}
\caption{Spatial distribution of the RR Lyrae stars found in this survey. 
RR Lyrae stars in the range of distances of $\omega$ Centauri are indicated with either black or red dots. For the later
we have measured radial velocities. Other RR Lyrae stars in the field (in either the foreground or background) are plotted with small star 
symbols. The black dashed circle corresponds to the tidal radius of $r_t=57$ arcmin of $\omega$ Centauri.
The solid square and dashed rectangle show the regions explored by \citet{leon2000} and \citet{dac08} respectively.}
\label{distribution}
\end{center}
\end{figure}

Figure ~\ref{distribution} shows the spatial distribution of the RR Lyrae stars detected in this work and the 25 stars within the distance limits above are marked with distinctive symbols. 
The distribution of those stars roughly along a line, as seen in this Figure,
should not be immediately interpreted as a tidal tail since it is just a reflection of the shape of the survey?s footprint (see Figure 1). Anyway, there are candidate stars to 
debris up to $\gtrsim 6$ degrees from the center of the cluster.\\

\section{Analysis}
\label{Sec:Analysis}

In the last section we identified a group of 15 RR Lyrae stars that are located in the same distance range as $\omega$ Centauri and hence, are potentially
tidal debris from the cluster. In this section we analyze this scenario from different perspectives.\\

\subsection{Comparison with the properties of the cluster RR Lyrae stars \label{properties}}

The sample of 15 candidates to debris is composed of 5 stars of the type $ab$ and 10 of the type $c$.
It is somewhat surprising that the number of type c star candidates is larger than the number of type ab (a ratio of 2), since in general, 
type c stars are more rare. In $\omega$ Centauri itself there are 76 RR$ab$ and 59 RR$c$ \citep{clement01}, a ratio of 0.77. Thus, the sample of the
candidates to cluster debris do not hold the same ratio. Miss-classification of the type c stars is a non-negligible possibility since, in the 
case of relatively few epochs in the light curves, eclipsing binaries of the type W UMa may mimic the
shape of type $c$ RR Lyrae stars. The contamination due to this type of stars is higher at lower galactic latitude, like in this case,  
where the disk population is important \citep[see][for a complete discussion]{vivas2004,mat2009}. High amplitude $\delta$ Scuti stars, which are also more common in the disk 
population, may have periods as long as type c stars and hence, they constitute an additional source of contamination. Hence, the true number of candidates to debris is likely lower than 15. \\

\begin{center}
\begin{figure}
\includegraphics[width=90mm]{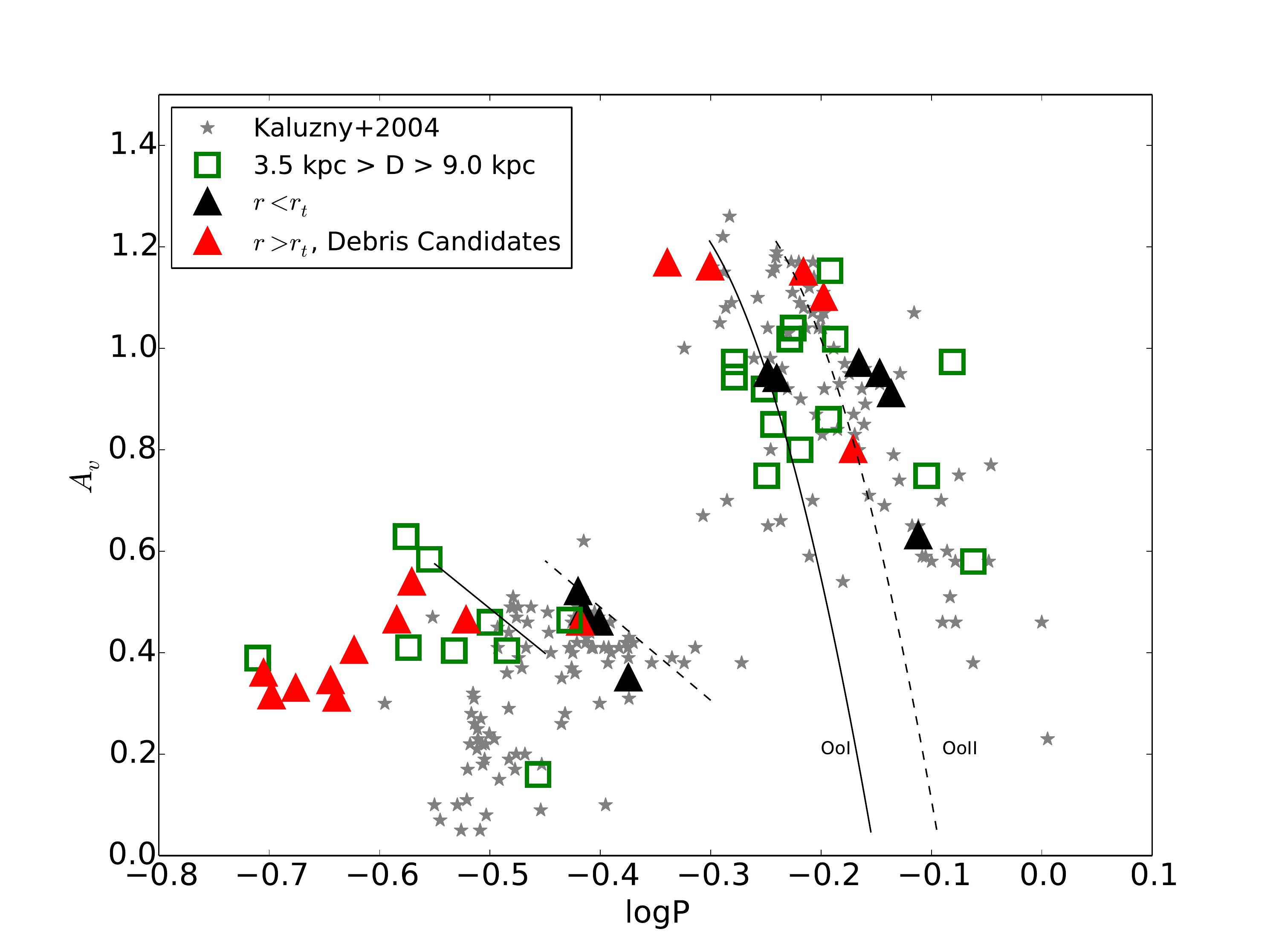}
 \caption{  Distribution of the RR Lyrae stars in the range of distances from 3.5 to 9 kpc in the Bailey (period-amplitude) diagram. 
 Stars within the tidal radius of the cluster are marked with black triangles while the ones outside the tidal radius (the candidates to debris)
 are indicated with red triangles.
 For reference, RR Lyrae stars in $\omega$ Centauri from \citet{kaluzny2004} are shown with star symbols. 
 Solid and dashed lines represent the typical locus of OoI  and OoII clusters, respectively \citet{Cacciari+2005}.}
 \label{Oosterhoff}
 \end{figure}
\end{center}

On the other hand, the mean period of the 5 type ab among the candidates to debris is 0.58 day. This is significantly different to the mean values of the periods
of type ab stars found by \citet[][86 stars, 0.656 days]{kaluzny2004} and \citet[][40 stars, 0.647 days]{weldra2007} in $\omega$ Centauri. The cluster has been given
an Oosterhoff II classification. Figure \ref{Oosterhoff} shows the distribution in the Bailey (period-amplitude) diagram for 
all debris candidates. For comparison, the diagram also contains the sample of RR Lyrae identified by \citet{Kaluzny+1997} in $\omega$ Cen. For the RR Lyrae stars in this work having 
only I magnitude, we used the relationship ${\rm Amp}_V = 0.075 + 1.497\times{}{\rm Amp}_I$ derived by \citet{Dorfi+1999}. 
Out of the 5 type ab candidates to debris (red triangles in Figure~\ref{Oosterhoff}), only 3 stars  have periods $>0.6$ day and lie within the OoII population locus.
The cluster RR Lyrae stars (star symbols) do show some dispersion in the Bailey's diagram and thus, the distribution of our candidates is not inconsistent with the properties
of the RR Lyrae stars in this diagram. \\

\subsection{Number of RR Lyrae stars expected in our survey}

Does this group of 15 RR Lyrae stars constitute an overdensity of stars over the expected population
of the halo/disk? In order to investigate this point, 
we calculated the expected number of RR Lyrae stars for the thick disc and halo over
the survey area and in the range of distance of our debris candidates  (3.5 kpc $< D < $9 kpc).
For these calculations, we used the density profiles of type ab RR Lyrae stars in the Galactic thick disc and halo compiled in Table 4 in \citet{mat2009}, 
which we reproduce here:\\

For the Halo, we used the \citet{preston91} model which has variable flattening density contours:

\begin{equation}\label{halo}
\rho_{halo}=\rho_{\odot}^{RR}\left(\frac{1}{R}\sqrt{x^2+y^2+\left(\frac{z}{c/a}\right)^2}\right)^n
\end{equation}

\noindent
where $(c/a)=0.5+(1-0.5(a/20))$ \citep[\textit{a} and \textit{c} in kpc,][]{preston91}.
The values for the slope and
local density were taken from \citet{vivas2006}, are $n=-3.1\pm0.1$ and 
$\rho_{\odot}^{RR}=4.2^{+0.5}_{-0.4}$ kpc$^{-3}$.\\

For the Thick Disk the number density is given by

\begin{equation}\label{disc}
 \rho_{disc}=\rho_{\odot}^{RR}exp\left(-\frac{R_{gal}-R_{\odot}}{h_R}\right)exp\left(-\frac{\mid z\mid}{h_z}\right)
\end{equation}

\noindent
where $h_R=0.51$ kpc, $h_z=2.20$ kpc \citep{Carrollo2010}. The local density of the thick disk
($\rho_{\odot}^{RR}=10$ kpc$^{-3}$)
was taken from \citet{layden1995}.\\

For the equations above we assumed the distance from the Sun to the galactic center as $R_{\odot}=8$ kpc \citep{Reid1993}. The total number of
RR Lyrae stars in each galactic component is given by integrating the above equations over the area of our survey and in the distance range of our candidates
to debris:

\begin{equation}
N_{RRLS}^{halo}=\int_{3.5 \mbox{ } kpc}^{9.0 \mbox{ } kpc} \int_{b} \int_{l}  \rho_{halo}(D,l,b)D^2\mbox{ }dD \mbox{ }cos(b)\mbox{ } db \mbox{ } dl  
\end{equation}

\begin{equation}
N_{RRLS}^{disc}=\int_{3.5 \mbox{ } kpc}^{9.0 \mbox{ } kpc} \int_{b} \int_{l}  \rho_{disc}(D,l,b)D^2\mbox{ }dD\mbox{ } cos(b)\mbox{ } db \mbox{ } dl
\end{equation}

Finally, the expected number of RR Lyrae stars ($ab$ type) in our survey area is:\\

\begin{equation}
 N_{RRLS}^{survey}=N_{RRLS}^{halo} + N_{RRLS}^{disc}
\end{equation}\\

Following the same method described in \citet{mat2009} to integrate those equations, we found
$N_{RRLS}^{halo}=(14.2\pm3.8)$ and $N_{RRLS}^{disc}=(3.7\pm1.9)$, for a total of $N_{RRLS}^{survey}=(17.9\pm4.2)$. If we assume a ratio $N_{ab+c}/N_{ab} = 1.29$ \citep{layden1995},
there should be $(23.1\pm4.8)$ RR Lyrae stars in the area of our survey.
Taking into account the different completeness level of our survey (which depend on the number
of epochs, see \S 3.2), we should expect ($22.2\pm4.7$) RR Lyrae stars in the survey. 
The cited errors are Poisson statistics.\\

We also explored the expected number of RR Lyrae stars using the recent characterization of the thick 
disk and halo by \citet{Robin2014}. 
These models predict $N_{RRLS}^{halo}=(18.5\pm4.3)$ and $N_{RRLS}^{disc}=(4.9\pm2.2)$ RR Lyrae stars (after completeness correction) 
depending on the thick disk model assumed 
\citep[Eq. 1 and 2 in][]{Robin2014}, in agreement with the previous estimates.\\

It is clear that the expected number of RR Lyrae stars in the galactic components is of the same 
order (or even slightly higher) to what we actually found in our survey.
Again, this does not favor the scenario of a significant amount of debris around the cluster since no
overdensity of RR Lyrae stars has been observed.\\

\subsection{Radial Velocities}

Any recent debris from the cluster would be expected to have a similar radial velocity to that of the cluster.
 In order to investigate this issue we were able to obtain eight low resolution spectra of stars in our survey.
Although the number is small, it can give us an idea if there is a preferential velocity among the candidates to debris.\\

\begin{center}
\begin{figure}
 \includegraphics[width=100mm]{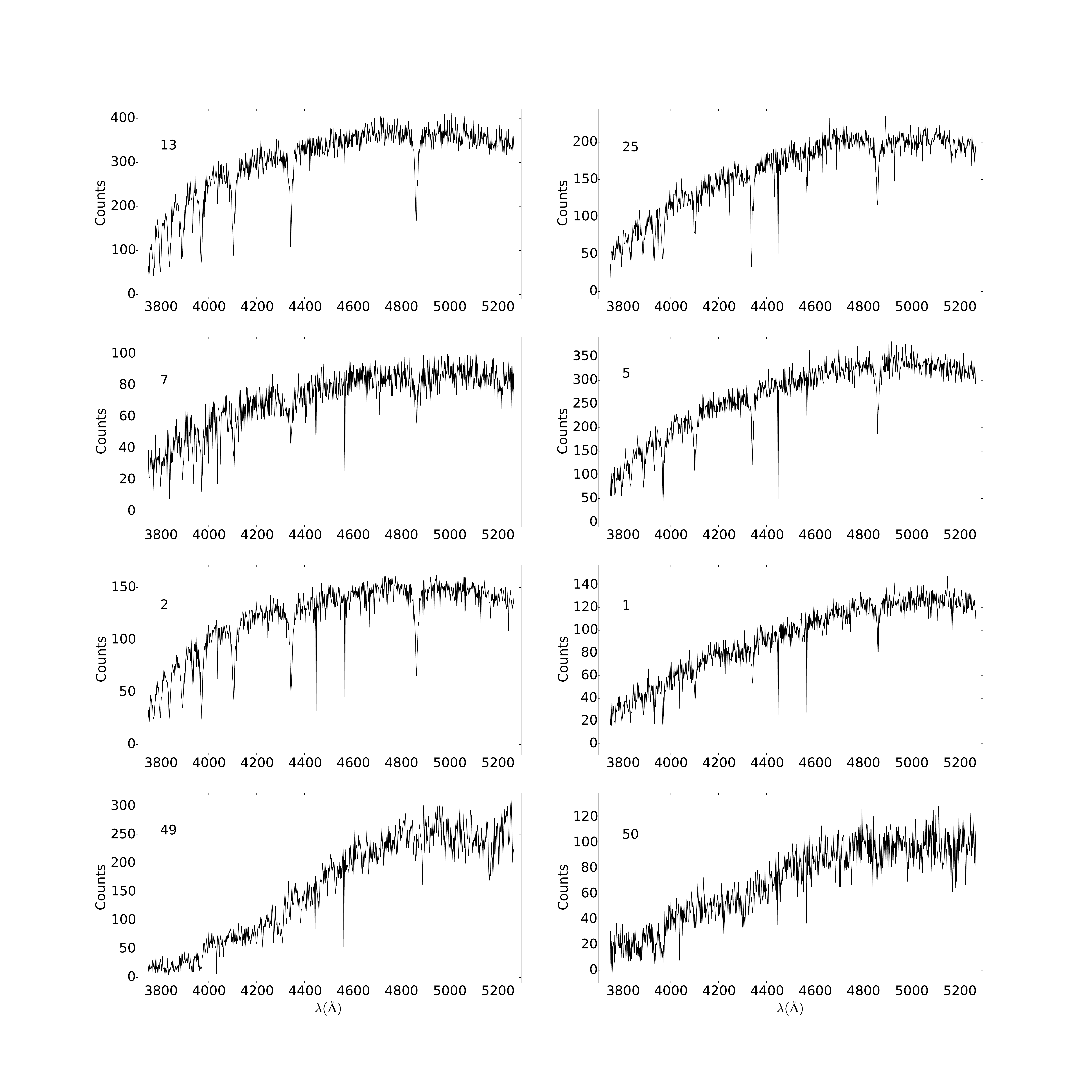}
 \caption{Spectra of the 8 variable stars observed with the SMARTS R-C spectrograph.}
 \label{spectral}
 \end{figure}
\end{center}

The spectra was obtained with the R-C Spectrograph at the 1.5m Telescope operated by the SMARTS consortium at Cerro Tololo 
Interamerican Observatory (CTIO), Chile, and were reduced using standard IRAF routines. 
We obtained a signal to noise ratio $S/N>30$ with exposure times of 900s for each of the 8 observed stars (Figure~\ref{spectral}). Some of the stars were
observed at two different epochs (different nights). 
Although we took spectra of 8 stars, 2 of them, which were initially classified as type c, showed spectra that were too cool for this type of stars.
Type c stars are in the bluest end of the instability strip region and hence have spectral types of A stars. The spectra of these 2 stars were 
late F/early G (see stars \#49 and \#50 in Figure~\ref{spectral}). This finding supports our claim (\S~\ref{properties}) that the sample of type c stars may be contaminated by other types of
variables. Coordinates and other data for these two stars are given in Table~\ref{New}.\\

\begin{table}
\setlength{\tabcolsep}{0.95mm}  
\centering
\begin{minipage}{100mm}
\begin{tiny}
\caption{Photometric Parameters for Stars \#49 and \#50}
\label{New}
\begin{tabular}{@{}cccccccc@{}}
\hline
ID & $\alpha_{2000}$       & $\delta_{2000}$      & $N_{obs}$  & P             & Amp     &  $\langle V \rangle$   &  $\langle I \rangle$  \\
   &  (deg)                &  (deg)               &            &       (day)   & (mag)   &     (mag)              & (mag)                 \\

\hline
49 & 202.643875 & -45.994461 & 17 & 0.41530 & 0.25  & 14.02$\pm$0.04 & -              \\
50 & 206.151932 & -44.192210 & 22 & 0.32759 & 0.19  & -              & 14.18$\pm$0.02  \\
\hline
\end{tabular}
\end{tiny}
\end{minipage}
\end{table}

\begin{center}
\begin{figure}
 \includegraphics[width=97mm,height=90mm]{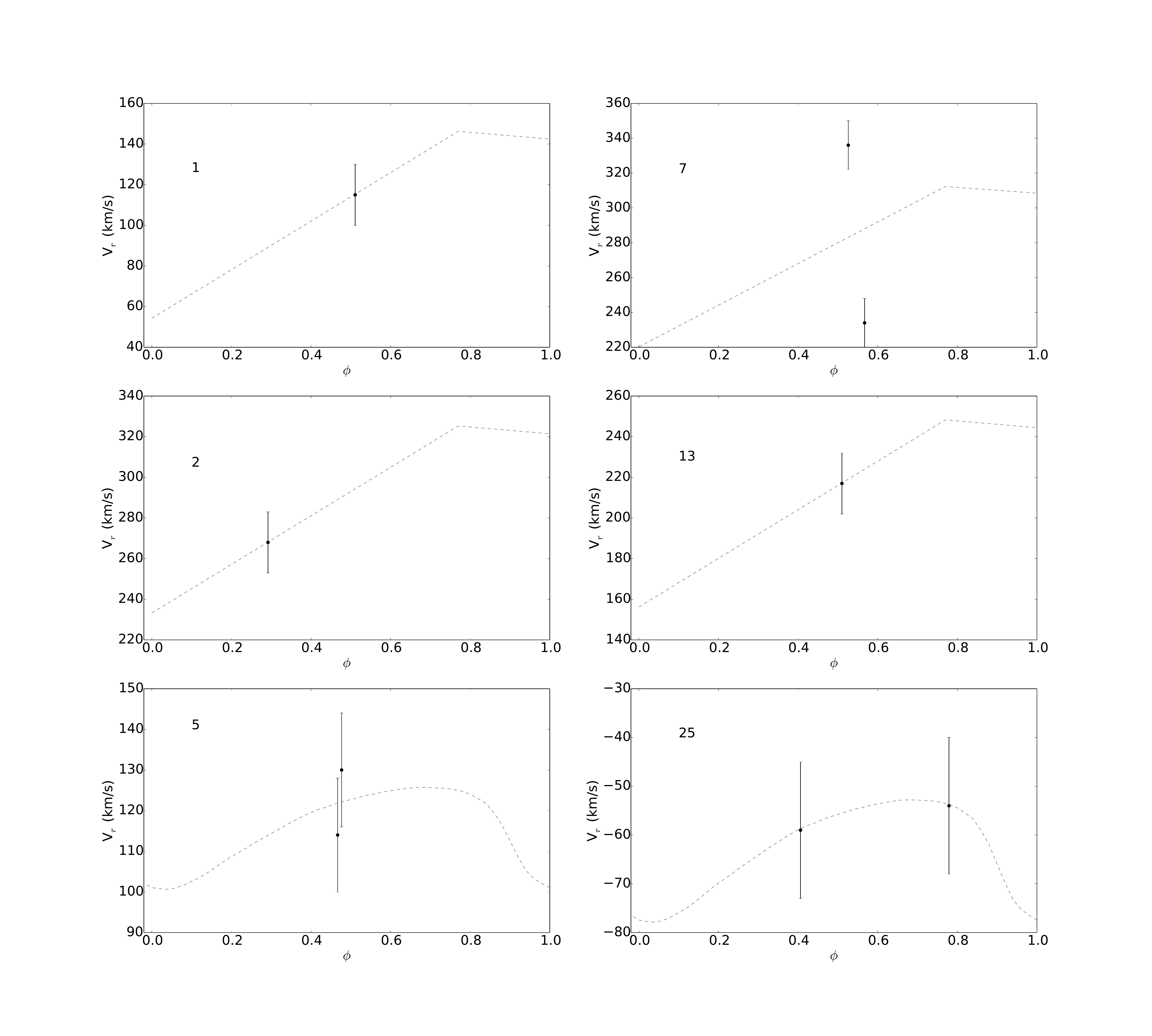}
 \caption{{\bf Two top rows:} Radial velocity curve template of X Ari (dashed line) for 
 RR Lyrae star $ab$-type. {\bf Bottom row:} Radial velocity template T Sex and DH Peg (dashed line) for 
 RR Lyrae star c-type (from \citet{Duffau06,Duffau14}).  
 The filled circles are the measured radial velocities.}
 \label{radialvelocity}
 \end{figure}
\end{center}

Radial velocities were then measured for 6 RR Lyrae stars via cross-correlation techniques using 
the IRAF routine $fxcor$. The templates for the correlation were five bright radial velocity
standards from a list compiled originally by \citet{layden1994}, which were observed with the same instrumental
setup as the RR Lyrae stars. The correlation was carried out over the wavelength range $\lambda \lambda$3700-5300 \AA, a 
spectral region that encompasses a number of strong features
such as the Ca II K-line and H-line, and the H$_{\delta}$, H$_{\gamma}$ and H$_{\beta}$ hydrogen lines.
Heliocentric corrections were then applied to correct for 
the Earth's motion. Systemic velocities were calculated by means of fitting a radial velocity curve template
\citep[see][for details]{vivas2008}. Figure ~\ref{radialvelocity} shows the templates used for obtaining the systemic
velocities together with the radial velocities measured for each star. Typical errors for the radial velocities are of the order of 20 km/s.
For star \#7, the template of the radial velocity curve does not agree well with the data. We are quoting an error for this star which is
the average difference between the observational points and the template.\\

\begin{figure}[ht!]
  \begin{center}
\centering
\includegraphics[width=92mm]{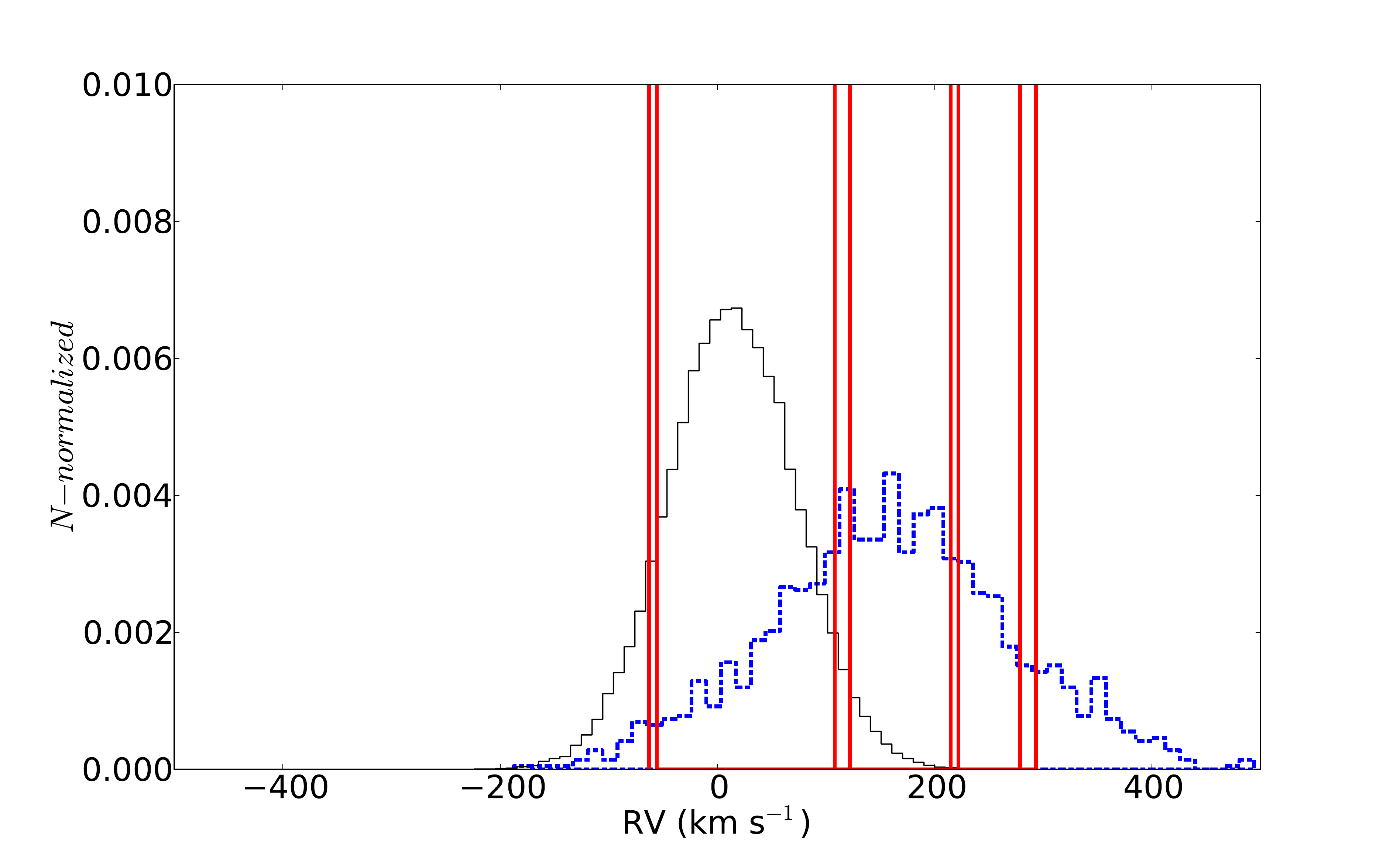}
 \caption{Radial velocity distributions of the halo (blue dashed histogram) and thick disk (black solid histogram) populations from the Besan\c{}con Galaxy Model \citep{Robin2003} in the
 area of the sky covered by our survey. 
 The red histogram correspond to the radial velocities of the six RR Lyrae stars 
 for which we obtained spectroscopic observations.}
 \label{distribution_velocity}
\end{center}
\end{figure}

In Figure~\ref{distribution_velocity} a histogram of the expected distribution of halo and thick disk stars in this part of the 
sky is plotted with a dashed and solid histogram, respectively. These distributions were 
obtained by extracting simulated
data out of the Besan\c{}con Galaxy model\footnote{\url{http://model.obs-besancon.fr}} \citep{Robin2003} in this part of the sky (using a simple selection function of stars with
$(V-I) \leq 0.95$ mag in colors and 14 mag $< V < 16$ mag).
The red histogram shows the location of the six RR Lyrae stars with radial velocity.
Of those six objects, stars \#7 and \#13 are both located within the tidal radius.
As expected they both have, within errors, similar radial velocity as the cluster 
\citep[232.5 km/s,][]{din99}. There are however no other stars that share the same velocity.
Thus, radial velocities do not provide any indication for debris around the cluster. The RR Lyrae stars have velocities consistent with them
belonging to the halo or thick disk population.\\

\subsection{Orbits}

 As a final test, we calculated the probability that each one of 
these 4 stars had a close encounter with the cluster sometime in the past.
If that is the case, it may be argued that the stars were stripped off the cluster a long time ago.
To do this, we calculated $10^5$ pairs of simulated orbits (for the cluster and each one of the RR Lyrae stars)
using an axisymmetric Milky Way-like galactic model, following the one by \citet{AllenSantillan1991}, scaled with the
values $R_0=8.3$ kpc and $\Theta_0 = 239$ km s$^{-1}$ \citep[see][]{brunthaler2011} and the methodology
described in \citet{Pichardo2012}. The axisymmetric background potential model consists of three
components, a Miyamoto-Nagai spherical bulge and disk and 
a supermassive spherical halo. In addition to the position, distance and radial velocity given in 
Table~\ref{table1}, the orbit calculation requires proper motions. These were obtained 
from the UCAC4 catalog \citep{zacha2013} and are given in Table ~\ref{Table3}.
For $\omega$ Centauri itself, we used the proper motions measured by \citet{din99}: $\mu_\alpha = -5.08\pm0.35$ mas;
$\mu_\delta \cos(\delta) = -3.57\pm0.34$ mas. \\

\begin{figure*}[ht!]
\begin{center}
 \includegraphics[width=137mm,height=95mm]{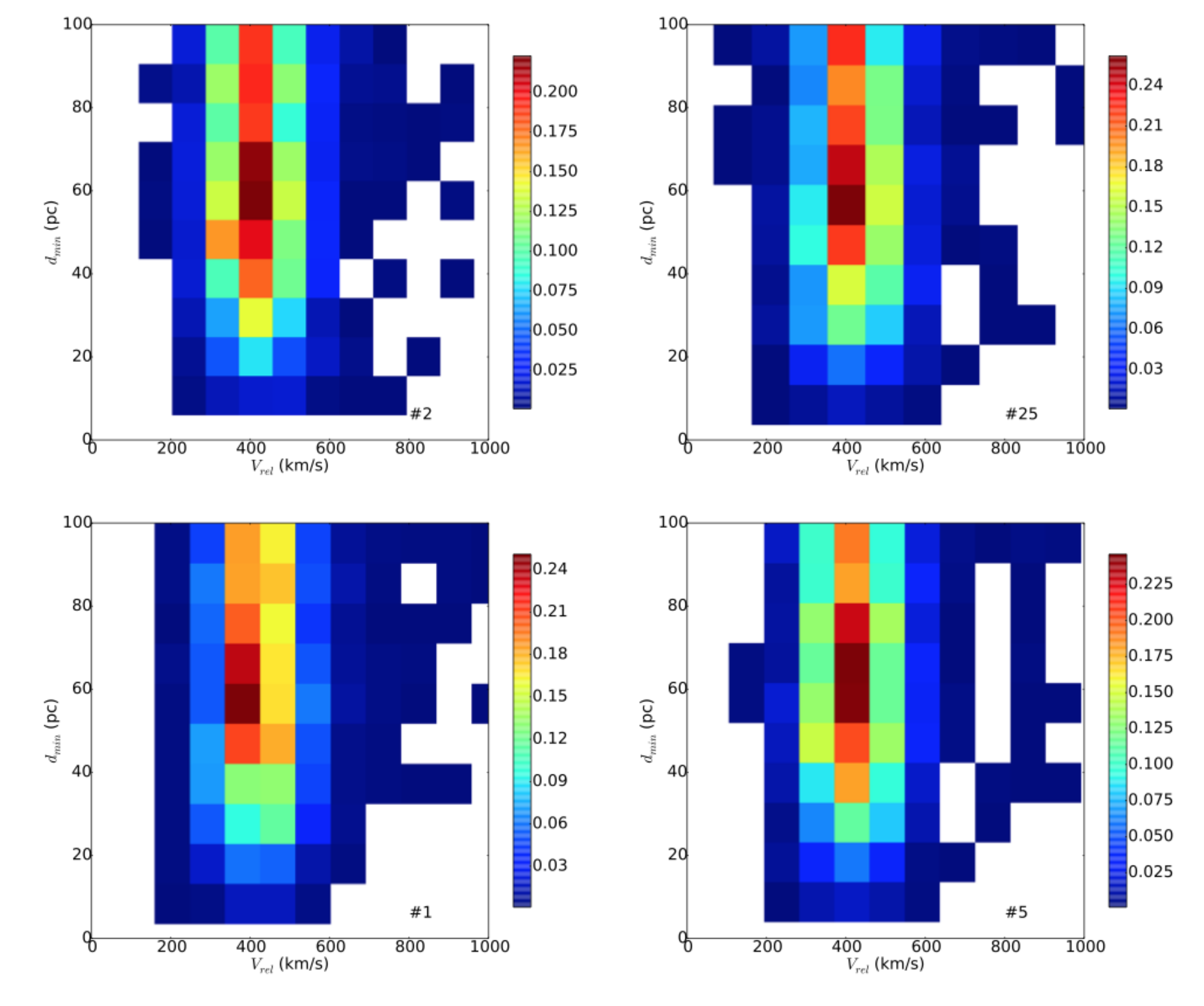}
 \caption{Distance of minimum approach ($d_{min}$) as a function of the relative star-cluster velocity 
 for close encounters within 1 Gyr in the orbit simulations. 
 The color code scales with the probability of a pair of orbits having such $d_{min}$ and $V_{rel}$.}
\label{simulations}
\end{center}
\end{figure*}

 We calculated the probability of a close encounter between the cluster and the 
stars, which was defined as having a minimum approach $d_{min}\leq100$ pc, or about the size of the
tidal radius of the cluster. Orbits were integrated up to 1 Gyr in the past. 
We found low probabilities for such close encounters between the stars and the cluster.
Even more, a closer examination of the circumstances of the few possible encounters casts doubts over the scenario 
of these stars to have been tidally stripped from the cluster.
Figure~\ref{simulations} shows the details of the close encounters for the 4 stars investigated here.
Each panel shows the distance of minimum approach ($d_{min}$) as a function of the relative velocity between the star
and cluster ($V_{rel}$) at the moment of the encounter. 
In all 4 cases $V_{rel}$ peaks at $\sim 400$ km/s. That implies the stars have a much higher velocity than the escape velocity at the cluster center and at the 
cluster half-mass radius  \citep[$V_0=60.4$ km s$^{-1}$ and $V_h=44$ km s$^{-1}$,][]{Gnedin2002}.
It is possible that a different mechanism such as ejection due to interaction with a binary
system may be responsible for this discrepancy in the velocities \citep[see for example][]{Hut1993, Pichardo2012}. 
Anyway, the relevant point related to our work is that 
tidal stripping would not be an adequate mechanism to explain the current position and velocity of those stars.\\

\begin{table}
\setlength{\tabcolsep}{5.mm}  
\centering
\begin{minipage}{100mm}
\begin{tiny}
  \caption{Input paramaters for the simulations.}
  \label{Table3}
 \begin{tabular}{@{}ccccc@{}}
\hline
ID                & $\alpha_{2000}$       & $\delta_{2000}$  &  $\mu_{\alpha}cos(\delta)$   & $\mu_{\delta}$ \\
                  &   (deg)   &   (deg)    &        (mas/yr)                &      (mas/yr)   \\
\hline
 1                & 199.787506 & -49.50225   &   8.5$\pm$2.8                &-8.4$\pm$2.8   \\
 2                & 199.855804 & -47.08439   &  -18.6$\pm$2.4               &3.3$\pm$2.6     \\
 5                & 200.793121 & -45.59691   &  -11.9$\pm$1.6               &-0.6$\pm$1.6   \\
 25               & 203.715485 & -45.77067   &  -11.1$\pm$2.9               &6.8$\pm$2.9    \\
\hline
\end{tabular}
\end{tiny}
\end{minipage}
\end{table}

\section{CONCLUSIONS}
\label{Sec:Conclusions}

We searched for tidal debris of the alleged progenitor galaxy of $\omega$ Centauri by using RR Lyrae
stars as tracers of its population. We found  48 RR Lyrae variables (25 RR$ab$ and 23 RR$c$) in a region
of $\sim 50$ deg$^2$. around the cluster.\\

Although several of those stars have a similar distance as the cluster, we conclude that 
there are no signs of stellar streams in the neighborhood of the globular cluster $\omega$ Centauri. This conclusion is based on the following:\\

\begin{itemize}

 \item      The expected number of RR Lyrae stars due to the halo and thick disk population of the Galaxy
            is consistent with the total number of RR Lyrae stars detected in this work. Thus, there is no 
            overdensity of these stars in this 
            region of the sky that can be linked to stellar debris.\\
 
 \item      None of the radial velocities we obtained for stars outside the cluster had a 
            radial velocity similar to that of the cluster.
            The caveat for this is that very few stars were measured spectroscopically.
            A firmer conclusion would necessarily need a more exhaustive spectroscopic study.\\
 
 \item      The orbit simulations for the RR Lyrae stars in our survey with radial velocity
            show a very low probability that the stars were torn off the cluster 
            in the past. The high relative velocity of the few possible close encounters
            suggests that the stars would have been ejected from the cluster by a different physical process
            other than tidal stripping.\\  
 
 \item      The ratio of between types $ab$ and $c$ among the stars at similar distance as the cluster
            is very different to the well known ratio of variables in the cluster. This suggests either it is a
            different stellar population and/or contamination for other type of stars is present in our survey. \\

\end{itemize}

This work confirms the results obtained by \citet{dac08}, who also did not find a significant amount 
of tidal debris around the cluster. It is still puzzling that debris material have been found
far away from the cluster (e.g. in the solar neighborhood) but not near the cluster itself.
Our survey covers a large area of sky, but it is not uniform around the cluster in the sense that
we preferentially surveyed along the path of the orbit of the cluster. We plan to pursue a larger 
and more uniform survey of RR Lyrae stars in the near future and include regions near the cluster 
not explored in this work.\\

\section*{Acknowledgments}
This research was based on observations collected at the J\"urgen Stock 1m Schmidt telescope and the 
1m Reflector telescope of the National Observatory of Llano del Hato Venezuela, 
which is operated by CIDA
for the Ministerio del Poder Popular para la Ciencia y Tecnolog{\'\i}a, Venezuela. 
J.G.F-T. acknowledges the support from Centre national d'\'etudes spatiale (CNES) 
through Phd grant 0101973 and UTINAM Institute of the Universit\'e de Franche-Comte, 
supported by the Region de Franche-Comte and
Institut des Sciences de l'Univers (INSU).
C.M. acknowledges the support of the post-doctoral fellowship of DGAPA-UNAM, M\'exico. R.Z. acknowledges the support of NSF grant AST11-08948.
This research was made possible through the use of the AAVSO Photometric All-Sky 
Survey (APASS), funded by the Robert Martin Ayers Sciences Fund.
We thank C. Navarrete for helpful discussions on some of the individual variables. We thank the anonymous referee for 
useful comments and suggestions.

{}


\end{document}